\documentclass[apj]{emulateapj}

\begin{document}                  

\title{Discovery of the Dust-Enshrouded Progenitor of SN~2008S with
{\it Spitzer}}

\author{Jos{\'e}~L.~Prieto\altaffilmark{2,4},
Matthew~D.~Kistler\altaffilmark{3,4},
Todd~A.~Thompson\altaffilmark{2,4}, Hasan~Y{\"u}ksel\altaffilmark{3,4},
Christopher~S.~Kochanek\altaffilmark{2,4},
Krzysztof~Z.~Stanek\altaffilmark{2,4},
John~F.~Beacom\altaffilmark{2,3,4},
Paul~Martini\altaffilmark{2,4}, Anna~Pasquali\altaffilmark{5}, and
Jill~Bechtold\altaffilmark{6}}

\altaffiltext{1}{Based in part on data acquired using the Large
Binocular Telescope (LBT). The LBT is an international collaboration
among institutions in the United States, Italy and Germany. LBT
Corporation partners are: The University of Arizona on behalf of the
Arizona university system; Istituto Nazionale di Astrofisica, Italy; LBT
Beteiligungsgesellschaft, Germany, representing the Max-Planck Society,
the Astrophysical Institute Potsdam, and Heidelberg University; The Ohio
State University, and The Research Corporation, on behalf of The
University of Notre Dame, University of Minnesota and University of
Virginia.}

\altaffiltext{2}{Dept.\ of Astronomy, The Ohio State University, 140
W.\ 18th Ave., Columbus, OH 43210; prieto, thompson, ckochanek,
kstanek@astronomy.ohio-state.edu}

\altaffiltext{3}{Dept.\ of Physics, The Ohio State University, 191 W.\
Woodruff Ave., Columbus, OH 43210; kistler, yuksel,
beacom@mps.ohio-state.edu}

\altaffiltext{4}{Center for Cosmology and AstroParticle Physics, 
The Ohio State University, 191 W.\ Woodruff Ave., Columbus, OH 43210}

\altaffiltext{5}{Max-Planck-Institut f\"ur Astronomie, K\"onigstuhl 17,
D-69117 Heidelberg, Germany; pasquali@mpia.de}

\altaffiltext{6}{Steward Observatory, University of Arizona, 933 North
Cherry Ave., Tucson AZ 85721-0065; jbechtold@as.arizona.edu}

\shorttitle{Mid-IR Detection of the SN~2008S Progenitor}

\begin{abstract}
We report the discovery of the progenitor of the recent type~IIn
SN~2008S in the nearby galaxy NGC~6946. Surprisingly, it was not found
in deep, pre-explosion optical images of its host galaxy taken with the
Large Binocular Telescope, but only through examination of archival {\it
Spitzer} mid-IR data.  A source coincident with the SN~2008S position is
clearly detected in the 4.5, 5.8, and 8.0~$\mu$m IRAC bands, showing no
evident variability in the three years prior to the explosion, yet is
undetected at 3.6 and 24~$\mu$m. The distinct presence of $\sim 440$~K
dust, along with stringent LBT limits on the optical fluxes, suggests
that the progenitor of SN~2008S was engulfed in a shroud of its own
dust. The inferred luminosity of $\approx 3.5\times10^4$~L$_\odot$
implies a modest mass of $\sim 10$~M$_\odot$. We conclude that objects
like SN~2008S are not exclusively associated with the deaths or
outbursts of very massive $\eta$ Carinae-like objects. This conclusion
holds based solely on the optical flux limits even if our identification
of the progenitor with the mid-IR source is incorrect.
\end{abstract}


%%%%%%%%%%%%%%%%%%%%%%%%%%%%%%%%%%%%%%%%%%%%%%%%%%%%%%%%%%%%%%%%%%%%%%%%%%%
%%%%%%%%%%%%%%%%%%%%%%%%%%%%%%%%%%%%%%%%%%%%%%%%%%%%%%%%%%%%%%%%%%%%%%%%%%%

\section{Introduction}
\label{sec:intro}

Over the last $\sim\,$20~years, several significant milestones have been
reached in the pre-explosion detection of core-collapse supernova
progenitors. These began with the ``peculiar'' type~II-P supernova 1987A
in the Large Magellanic Cloud (e.g., Menzies et al. 1987), where a
cataloged $\sim 20$~${\rm M}_{\odot}$ blue supergiant star was
identified as the progenitor (Sk $-$69 202; e.g., West et al. 1987).
Next came the transition type~IIb 1993J in M81, with a progenitor
identified as a red supergiant in a binary system (e.g., Podsiadlowski
et al. 1993; Maund et al. 2004). During the last decade, analyses of
pre-explosion archival optical imaging of nearby galaxies obtained
(mainly) with the {\it Hubble Space Telescope} have convincingly shown
red supergiants with masses $8 \, {\rm M}_{\odot} \leq {\rm M} \leq
20$~${\rm M}_{\odot}$ to be the typical progenitors of type~II-P
supernovae (e.g., Smartt et al. 2004; Li et al. 2007), the most common
core-collapse supernovae. Curiously, the progenitors of nearby type~Ib/c
supernovae, thought to result from very massive ($\ga 20$~${\rm
M}_{\odot}$) stars with strong winds that end their lives as Wolf-Rayet
stars, have evaded optical detection (e.g., Crockett et al. 2008).

The rarest and most diverse class of core-collapse supernovae are the
type~IIn (Schlegel 1990), which represent $\sim 2-5\%$ of all type~II
supernovae (e.g., Capellaro et al. 1997).  Their optical spectra,
dominated by Balmer lines in emission, and slowly declining light curves
show clear signatures of interactions between the supernova ejecta and a
dense, hydrogen-rich circumstellar medium (e.g., Filippenko
1997). Mainly due to their low frequencies, high mass loss rates, and
the massive circumstellar envelopes generally required to explain the
observations, some luminous type~IIn supernovae have been associated
with the deaths of the most massive stars (e.g., Gal-Yam et al. 2007;
Smith 2008 and references therein). Recently, evidence for this
association has increased with the report of a very luminous source in
pre-explosion images of the type~IIn SN~2005gl (Gal-Yam et al. 2007) and
the discovery of an LBV eruption two years before the explosion of
SN~2006jc (Pastorello et al. 2007). On the other hand, some low
luminosity type~IIn have been associated with the super-outbursts of
LBVs like $\eta$ Carinae (e.g., Van Dyk et al. 2000; Van Dyk et
al. 2006).

The appearance of the type~IIn SN~2008S in the nearby galaxy NGC~6946
($d\simeq 5.6$~Mpc; Sahu et al. 2006) was fortuitous, since a massive
stellar progenitor would be relatively easy to find. However,
pre-explosion images serendipitously obtained from the Large Binocular
Telescope revealed nothing at the position of SN~2008S, allowing us to
put stringent limits on the optical emission. In this {\it Letter}, we
report the discovery of an infrared point source coincident with the
site of SN~2008S using archival {\it Spitzer} Space Telescope data. The
{\it Spitzer} mid-IR detection, and deep optical non-detections, of the
progenitor are the tell-tale signs of a $\sim 10$~M$_\odot$ star
obscured by dust. We describe the available data in \S~\ref{sec:data},
our analysis in \S~\ref{sec:shroud}, and our conclusions in
\S~\ref{sec:discuss}.

%%%%%%%%%%%%%%%%%%%%%%%%%%%%%%%%%%%%%%%%%%%%%%%%%%%%%%%%%%%%%%%%%%%%%%%%%%%

\section{Searching for the Progenitor}
\label{sec:data}

NGC~6946 is quite a remarkable galaxy, giving birth to (at least) nine
SNe in the last century. The latest event discovered in NGC~6946 is
SN~2008S, found on February 1.79 UT at $\sim 17.6$~mag (Arbour \& Boles
2008) and located 52$\arcsec$~West and 196$\arcsec$~South of the nucleus
of NGC~6946. It was spectroscopically classified as a likely young
type~IIn supernova from the presence of narrow Balmer lines in emission,
highly reddened by internal extinction with a measured Na~D absorption
equivalent width of 5~\AA\ (Stanishev et al. 2008). Steele et al. (2008)
later reported that it had a peculiar spectrum due to the presence of
narrow emission lines from the [Ca~II]~730~nm doublet, Ca~II infrared
triplet, and many weak Fe~II features. The spectral properties and low
peak luminosity led Steele et al. (2008) propose that SN~2008S was a
{\em supernova impostor} such as SN~1997bs (Van Dyk et al. 2000).

Accurate coordinates are needed in order to search for the progenitor in
pre-explosion images. Fortunately, {\it Swift} started monitoring
SN~2008S with UVOT and XRT shortly after the discovery. We retrieved the
UVOT $ubv$ optical images obtained on Feb. 4.8, 6.0, and 10.5 (UT) from
the {\it Swift} archive.  We used WCSTools v3.6.7 (Mink 1999) and the
USNO-B astrometric catalog (Monet et al. 2003) to obtain astrometric
solutions for the images. The mean coordinates of SN~2008S are $\alpha =
20^{\rm h}34^{\rm m}45\fs 37$, $\delta = 60^{\circ} 05' 58.3''$
(J2000.0), with rms uncertainties of $\sigma_{\alpha}=0\farcs 5$ and
$\sigma_{\delta}=0\farcs 3$.

%%%%%%%%%%%%%%%%%%%%%%%%%
\begin{figure}[t]
\begin{center}
\includegraphics[width=3.25in,clip=true]{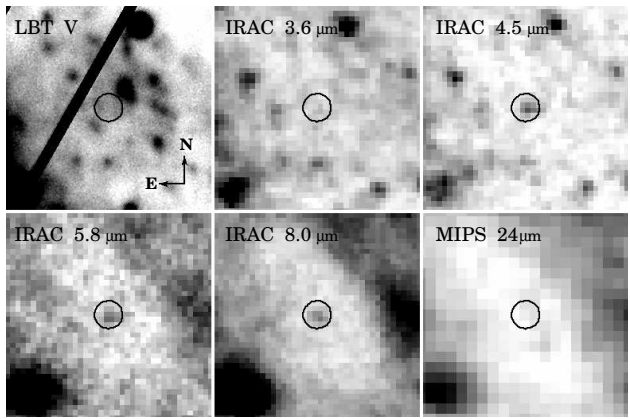}
\caption{Pre-supernova images ($30\arcsec \times 30 \arcsec$) of the
site of SN~2008S.  We show the LBT/LBC optical non-detection of the
progenitor and the images obtained with {\it Spitzer} by the SINGS
project at 3.6, 4.5, 5.8, 8.0, and 24~$\mu$m.  The progenitor is clearly
detected at 4.5, 5.8, and 8.0~$\mu$m.  The circle in each panel has a
radius of 2$\arcsec$ and is centered on the position of the supernova,
corresponding to 4 times the astrometric uncertainty of 0$\farcs$5. The
dark line in the LBT image is bleeding from a saturated star.}
\label{mosaic}
\end{center}
\end{figure}
%%%%%%%%%%%%%%%%%%%%%%%%% 

The Large Binocular Telescope (Hill et al. 2006) obtained deep optical
images of NGC~6946 on 19$-$21 May 2007, 225~days before discovery,
during Science Demonstration Time using the LBC/Blue camera (Ragazzoni
et al. 2006; Giallongo et al. 2008).  We combined the $12\times 300$~sec
images obtained using the $U$ filter (seeing $1\farcs 0$), and the
$4\times 300$~sec images obtained using the $B$ and $V$ filters (seeing
$1\farcs 5$).  After finding an astrometric solution for the combined
images using the USNO-B catalog ($\sigma_{\alpha}\simeq \sigma_{\delta}
= 0\farcs 2$), we do not detect a source at the position of SN~2008S
(see Fig.~\ref{mosaic}).  After calibrating the images using ancillary
optical data obtained by the Spitzer Infrared Nearby Galaxies Survey
(SINGS; Kennicutt et al. 2003) and {\it Swift}, we obtain 3$\sigma$
upper limits on the progenitor magnitudes of $U> 25.8$, $B> 25.9$ and
$V> 26.0$, which correspond to absolute magnitudes $M_{U} > -4.8$,
$M_{B} > -4.3$, and $M_{V} > -3.8$, correcting for $A_{V} = 1.1$~mag of
Galactic extinction (Schlegel et al. 1998). The upper limits are
calculated using aperture photometry from the standard deviation of the
sky at the SN position using a 10~pixel ($2\farcs 2$) diameter
aperture. We correct these values with aperture corrections estimated
using bright stars. The $\sim 0.2$~mag uncertainties in the 3$\sigma$
upper limits are due to the uncertainties in the aperture corrections
and the standard deviation of the sky (which is estimated from the rms
variations in the standard deviation measured in equal-sized apertures
placed in the background around the SN position). Welch et al. (2008)
reported 3$\sigma$ upper limits from pre-explosion Gemini/GMOS
observations of $V>24.0$, $R>24.5$ and $I>22.9$. These correspond to
absolute magnitudes $M_{V} > -6.8$, $M_{R} > -5.2$, and $M_{I} > -6.5$,
correcting for Galactic extinction.

Such a deep non-detection led us to investigate IRAC ($3.6-8.0$~$\mu$m;
Fazio et al. 2004) and MIPS ($24-160$~$\mu$m; Rieke et al. 2004) images
obtained by the SINGS Legacy Survey in 2004. We astrometrically
calibrated the images in the same way as the optical images from {\it
Swift} and LBT.  We detect a point source at $\alpha = 20^{\rm h}34^{\rm
m}45\fs 35$, $\delta = 60^{\circ} 05' 58.0''$ in the 4.5, 5.8, and
8.0~$\mu$m IRAC bands (see Fig.~\ref{mosaic}), with rms uncertainties
$\sigma_{\alpha}=0\farcs 5$, $\sigma_{\delta}=0\farcs 2$.  This is
consistent with the position of SN~2008S given the estimated
uncertainties, and thus likely to be the progenitor. The source is not
detected at 3.6, 24, or 70~$\mu$m. We estimate a probability of random
coincidence given the uncertainty in the SN position ($0\farcs 5$) of
0.8\% (0.02\%) from the density of 4.5~micron sources (with
[3.6]$-$[4.5] $>$ 1.5~mag) detected within a $1\arcmin$ radius of the SN
position.

%%%%%%%%%%%%%%%%%%%%%%%%%%%%%%%%%%%%%%%%%%%%%%%%%%%%%%%%%%%%%%%%%%%%%%%%%%%%  
\begin{figure}[t]
\plotone{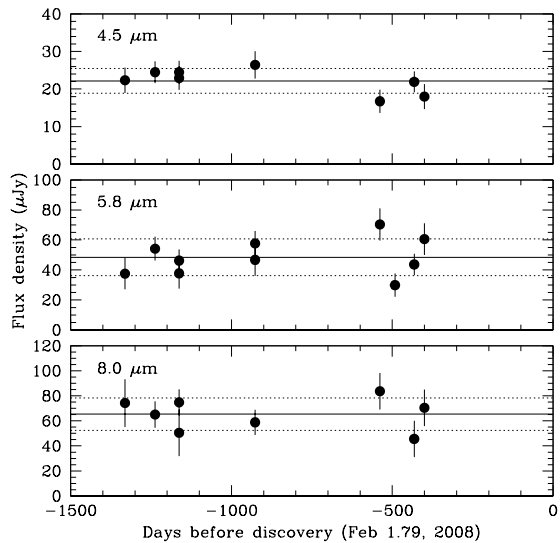}
\caption{Flux densities at 4.5, 5.8, and 8.0~$\mu$m as a function of
time (in days before the discovery) for the progenitor of SN~2008S.  The
solid line in each panel shows the mean for each band and the dashed
lines show the rms deviations of $\pm 3.3$, 12.2, and 13.0~$\mu$Jy,
respectively.}
\vspace{0.05in}
\label{spitzer}  
\end{figure}  
%%%%%%%%%%%%%%%%%%%%%%%%%%%%%%%%%%%%%%%%%%%%%%%%%%%%%%%%%%%%%%%%%%%%%%%%%%%%

We searched the {\it Spitzer} archive for all the programs that have
observed NGC~6946. Observations by the SINGS survey (PID: 159), and two
programs (PIs: Meikle, Sugerman, Barlow) monitoring the type~II-P SNe
2002hh and 2004et (PID: 230, 20256, 30292, 30494) provide a 2.5-year
baseline (June 2004 $-$ January 2007) of IRAC and MIPS observations
prior to the discovery of SN~2008S. We used aperture photometry (a
2~pixel extraction radius with aperture corrections) in the
flux-calibrated images provided by the {\it Spitzer Science Center} to
derive light curves for the progenitor.  Fig.~\ref{spitzer} shows the
flux density as a function of time in the 4.5, 5.8, and 8.0~$\mu$m bands
starting from June~2004.  There is no sign of variability at the $\sim
10 \%$ level. The non-detection at 3.6, 24, and 70~$\mu$m in single and
stacked images allows us to place useful upper limits on these fluxes.

Finally, we searched the {\it Chandra} archive to determine if the
progenitor was an X-ray source.  All five ACIS-S observations of
NGC~6946 include the location of SN~2008S.  These observations include a
60~ks exposure in 2001, a 30~ks exposure in 2002, and $3\times 30$~ks
exposures in 2004.  No source is detected at the supernova position in
any of these images. We set a 3$\sigma$ upper limit on the flux of the
progenitor of $f_X < 3 \times 10^{-15}$~erg~cm$^{-2}$~s$^{-1}$ ($L_{X} <
10^{37}$~erg~s$^{-1}$) in the broad X-ray band (0.5$-$8 keV), which
rules out a bright X-ray binary as the progenitor. This flux limit
corresponds to 20 counts in the longest exposure. Table~\ref{table:phot}
summarizes the detections and 3$\sigma$ upper limits on the progenitor
fluxes.

%%%%%%%%%%%%%%%%%%%%%%%%%%%%%%%%%%%%%%%%%%%%%%%%%%%%%%
\begin{table}[!t]
\begin{center}
\caption{Spectral Energy Distribution of the Progenitor of SN~2008S
\label{table:phot}}
\begin{tabular}{lccccc}
\hline \hline
\\
\multicolumn{1}{c}{} &
\multicolumn{1}{c}{$\lambda$} &
\multicolumn{1}{c}{} &
\multicolumn{1}{c}{$\lambda F_{\lambda}$} &
\multicolumn{1}{c}{Source} \\
\multicolumn{1}{c}{} &
\multicolumn{1}{c}{} &
\multicolumn{1}{c}{} &
\multicolumn{1}{c}{($10^{-17}\,\,{\rm W\,\,m^{-2}}$)} & 
\multicolumn{1}{c}{} \\
\\
\hline
\\
& 0.3-8~keV    &  & $< 0.3$ &  {\it Chandra}/ACIS-S \\
& 0.36~$\mu$m  &  & $< 0.07 $ & LBT/LBC-Blue \\
& 0.44~$\mu$m  &  & $< 0.11 $ & LBT/LBC-Blue \\
& 0.55~$\mu$m  &  & $< 0.08 $ & LBT/LBC-Blue \\
& 0.64~$\mu$m  &  & $< 0.22$  & Welch et al. (2008) \\
& 0.80~$\mu$m  &  & $< 0.63$  & Welch et al. (2008) \\
& 3.6~$\mu$m   &  & $< 0.45$ & {\it Spitzer}/IRAC \\ 	
& 4.5~$\mu$m   &  & $1.47\pm0.22$ &{\it Spitzer}/IRAC \\
& 5.8~$\mu$m   &  & $2.54\pm0.64$ &{\it Spitzer}/IRAC \\
& 8.0~$\mu$m   &  & $2.48\pm0.50$ &{\it Spitzer}/IRAC \\
& 24~$\mu$m    &  & $< 1.20$ & {\it Spitzer}/MIPS	\\	
& 70~$\mu$m    &  & $< 40$  & {\it Spitzer}/MIPS	\\	
\\
\hline
\hline
\end{tabular}
\end{center}
\end{table}
%%%%%%%%%%%%%%%%%%%%%%%%%%%%%%%%%%%%%%%%%%%%%%%%%%%%%%%%%%%%%%%%

\section{Beneath the Shroud}
\label{sec:shroud}

\begin{figure}[!ht]
\plotone{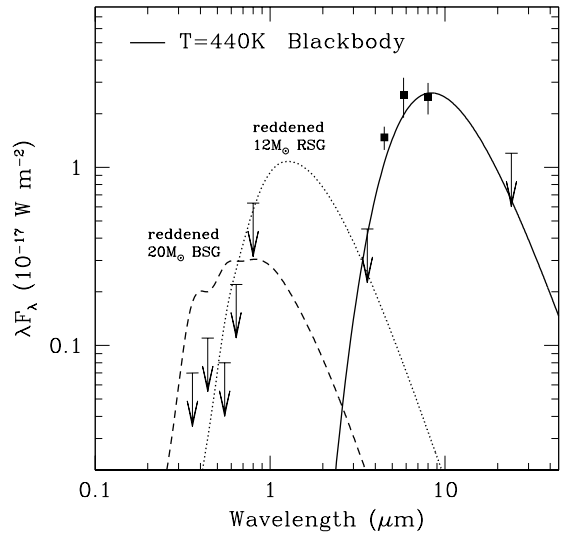}
\caption{The spectral energy distribution of the progenitor of SN~2008S
from {\it Spitzer} observations.  Detections are shown as open squares
at 4.5, 5.8, and 8.0~$\mu$m. Upper limits (3$\sigma$) from the combined
images at 3.6 and 24~$\mu$m are also indicated.  The solid line is the
best-fit blackbody with $T = 440$~K. We also show the $UBV$ upper limits
(3$\sigma$) from LBT and the $RI$ limits from Welch et al. (2008). The
measured fluxes are not extinction corrected. The dotted line shows a
reddened blackbody with the luminosity ($10^{4.5}$~L$_{\odot}$) and
effective temperature (10$^{3.5}$~K) of a 12~M$_{\odot}$ red
supergiant. The dashed line shows a reddened blackbody with the
approximate temperature (10$^{4.2}$~K) and luminosity
(10$^5$~L$_{\odot}$) of the blue supergiant progenitor of SN~1987A,
which has similar properties to the lowest luminosity LBVs observed
(e.g., Smith 2007). The models were reddened with $A_{V}=2.5$~mag, the
total extinction estimated from the colors of SN~2008S. \label{SED}}
\end{figure}

The measured fluxes and upper limits in the mid-IR bands are shown in
Fig.~\ref{SED}.  The shape of the spectral energy distribution (SED)
suggests thermally-radiating dust as the source of the emission. We
derive a best-fit single-temperature blackbody of $T\simeq 440$~K, with
a luminosity of $L_{\rm bol} \simeq 3.5 \times 10^4$~L$_\odot$
($d=5.6$~Mpc; Sahu et al. 2006), which implies a blackbody
radius\footnote{These values would change to $L_{\rm bol} \simeq 8
\times 10^4$~L$_\odot$ and $R_{\rm BB} \simeq 230$~AU if we assume an
extreme distance to NGC~6946 of 8.5~Mpc, which is the 3$\sigma$ upper
limit of the distance used by Li et al. (2005; 5.5$\pm$1.0~Mpc).}
$R_{\rm BB} \simeq 150$~AU. This luminosity points to a $\sim
10$~M$_{\odot}$ star at the end of its life (e.g., Meynet \& Maeder
2003). The 3$\sigma$ upper limit at 70~$\mu$m further limits the total
luminosity of the dust-enshrouded source and the geometry of obscuring
dust distribution.

%%%%%%%%%%%%%%%%%%%%%%%%%%%%%%%%%%%%%%%%%%%%%%%%%%%%%%%%%%%%
%%%%%%%%%%%%%%%%%%%%%%%%%%%%%%%%%%%%%%%%%%%%%%%%%%%%%%%%%%%%%%%

As shown in Fig.~\ref{SED}, a blackbody yields a relatively poor fit to
the data ($\chi^2 \approx 4.9$ per d.o.f.).  The inability of a
single-temperature blackbody to accommodate the data follows primarily
from the rapid change in the SED implied by the 3.6~$\mu$m upper limit
and the 4.5~$\mu$m detection.  Radiation transport calculations using
{\tt DUSTY} (Ivezic \& Elitzur 1997) were performed as a sanity check.
Using a central incident blackbody with $T = 3000 - 20000$~K we
calculated the emergent spectrum from a spherical dusty shell extended
over approximately one decade in radius.  As expected, the best
correspondence with the data is obtained for a total optical depth at
8.0~$\mu$m of order unity, although the precise value depends on the
assumed radial gradient of the density, the radial extent of the
obscuring medium, and the mixture of grain types.  Although a detailed
investigation of the dust properties is beyond the scope of this {\it
Letter}, we note that the strong evolution in the SED between 3.6 and
4.5~$\mu$m may signal the need for relatively large grains (e.g., Ivezic
\& Elitzur 1996).

We can estimate the mass of obscuring gas and dust by assuming that the
medium is marginally optically thick at 8.0~$\mu$m. Setting
$\tau_\lambda \approx 1\approx\kappa_\lambda\rho R_{\rm BB}$, and
assuming $\rho=M/(4\pi R_{\rm BB}^3/3)$, we find that $M \sim
10^{-3}\kappa_{\lambda,\,10}^{-1}$~M$_\odot$, where
$\kappa_{\lambda,\,10}=\kappa_\lambda/10$~cm$^2$~g$^{-1}$ is a typical
value for the Rosseland-mean dust opacity for gas at $\sim 440$~K (e.g.,
Semenov et al.~2003). This suggests a gas density on the scale $R_{\rm
BB}$ of $n \sim 3 \times 10^7\,\,{\rm cm^{-3}}$.  We also estimate a
minimum mass loss rate from the progenitor of $\dot{M}_{\rm min}\,=\,4
\pi R^2_{\rm BB}\rho c_g \sim 10^{-5}$~M$_{\odot}\,\,{\rm yr}^{-1}$,
where $c_g \sim 2$~km~s$^{-1}$ is the gas sound speed in the medium on
the scale of $R_{\rm BB}$.

The lack of variability in the mid-IR fluxes (see Fig.~\ref{spitzer})
limits the expansion velocity of the photosphere. Given our estimated
temperature and luminosity, keeping the mid-IR fluxes constant to within
$\sim 10$\% over the $\sim 10^3$~days covered by the observations means
that the dust photosphere cannot be expanding by more than $\sim
10$~km~s$^{-1}$, which is below the escape velocity of 13~km~s$^{-1}$
for a $10$~M$_\odot$ star at the estimated photospheric radius of
$150$~AU. This is further evidence that the dust is part of a relatively
steady, massive wind rather than an explosively-expelled dust shell.

\section{Discussion and Conclusions}
\label{sec:discuss}

Our pre-explosion detection of the progenitor of the type~IIn SN~2008S
is, to the best of our knowledge, the first in the mid-IR. The {\it
Spitzer} observations suggest an enshrouded star with a mass of $\sim
10$~M$_{\odot}$, buried in $\sim 10^{-3}$~M$_\odot$ of gas and dust. If
SN~2008S was a real supernova explosion, this is direct evidence that
relatively low-mass stars can end their lives as type~IIn SNe when they
have a sufficiently dense CSM from a massive wind, as proposed by Chugai
(1997). If this event was the luminous outburst of an LBV, it presents
evidence for low-luminosity, low-mass LBVs that have not been observed
before\footnote{The lowest-mass LBVs known have initial masses of
$20-25$~M$_{\odot}$ and luminosities $> 10^{5}$~L$_{\odot}$ (e.g., Smith
et al. 2004; Smith 2007).}. These conclusions about the relatively low
mass hold even if the identification of the progenitor with the Spitzer
source is incorrect. In this case, we know the total extinction from the
colours of the SN (see below). As shown in Fig.~\ref{SED}, our optical
limits with this extinction correspond to mass limits of $ \la
12$~M$_{\odot}$ for red supergiants and $ \la 20$~M$_{\odot}$ for blue
supergiants\footnote{We obtain an upper limit in the absolute optical
magnitude of the progenitor of $M_{V} \ga -7.1$ if we assume an upper
limit on the extinction estimate from the SN color ($A_{V}\simeq
3.5$~mag; bluest black-body possible) and an extreme distance to
NGC~6946 of 8.5~Mpc.}.

Interestingly, we see luminous dust-enshrouded stars in the Milky Way
and the LMC whose physical properties match well the observed properties
of the progenitor of SN~2008S. van~Loon et al. (2005, and references
therein) studied the properties ($T_{\star}$, $T_{\rm dust}$, $L_{\rm
bol}$, $\dot{M}$) of dust-enshrouded AGB stars and red supergiants in
the LMC using mid-IR observations.  These are M-type stars with
effective temperatures $\sim 2500-3800$~K, which have strong winds with
high (gas + dust) mass loss rates ($\dot{M} \sim 10^{-6} -
10^{-3}$~M$_\odot$~yr$^{-1}$), and warm dust emission from their dusty
envelopes (200~K $< T_{\rm dust} <$ 1300~K). Due to these similarities,
we conclude that the progenitor of SN~2008S was likely a dust-enshrouded
AGB (core-collapse produced from electron capture in the O-Ne-Mg core,
e.g., Eldridge et al. 2007; Poelarends et al. 2008) or red supergiant
like the ones observed in the LMC.

Although the detection and physical properties of the progenitor are the
main results of this study, we can also try to understand something
about the progenitor and explosion mechanism from the supernova itself.
The classification spectrum of SN~2008S is similar to the published
spectrum of SN~1997bs (Van Dyk et al. 2000), which showed narrow Balmer
lines in emission and many weaker Fe~II lines (V.~Stanishev,
priv. comm.; Steele et al. 2008).  SN~2003gm had photometric and
spectroscopic characteristics similar to SN~1997bs (Maund et al. 2006).
Since both of these were faint ($M_{V} \sim -14$~mag) compared with the
typical absolute magnitudes at maximum of type~II SNe ($M_{V} \sim -16$
to $-18$~mag), it is still debated whether they were intrinsically faint
explosions or super-outbursts of LBVs.

The early optical photometry obtained with {\it Swift} also indicates
that SN~2008S was a low-luminosity object, with $M_{V} \sim -14$~mag
after correcting for the total extinction along the line of sight. We
estimate the total extinction for $R_{V}=3.1$ to be $A_{V} \approx
2.5$~mag from the observed color $B-V \simeq 0.8$~mag and assuming a
typical intrinsic temperature of $\sim 10000$~K at this early phase of
the evolution. This value is roughly consistent with the estimated
reddening obtained from the reported equivalent width of the Na~D
absorption feature ($ 2.5 < A_{V} < 7.8$; based on Turatto et al. 2002).
This implies the presence of significant internal extinction with
$A_{V}\simeq 1.4$~mag after correcting for $A_{V}({\rm Gal}) =
1.1$~mag. Although the light from the supernova likely destroyed the
dust that obscured the progenitor to significantly beyond the blackbody
scale of $\sim 150$~AU, the existence of internal extinction in the
supernova light curve implies a more tenuous dusty obscuring medium on
larger scales. In fact, the rare detection of the [Ca II]~730~nm doublet
in emission by Steele et al. (2008) may provide direct, and independent,
evidence for a significant amount of dust in the CSM that was destroyed
by the UV-optical flash (e.g., Shields et al. 1999). The future spectra
and light curves of SN~2008S, optical as well as radio and X-ray, should
further probe the environment as they show signs of interactions with
the progenitors's wind.
 
The field of supernova forensics has advanced rapidly in recent years,
with $\sim 10$ SN progenitors now known (e.g., Smartt et al. 2004; Li et
al. 2007). Moving forward, several groups are obtaining the data
required to more fully characterize the progenitors of future nearby SNe
(e.g., Kochanek et al. 2008). We note that the discovery of the
progenitor of SN~2008S itself would not have been possible only a few
years ago without {\it Spitzer}. Future multi-wavelength surveys of the
local universe are thus encouraged in order to catch other unexpected
stellar phenomena, potentially even before they occur.

\acknowledgments

We thank B.~Blum, J.~Eldridge, M.~Elitzur, A.~Gal-Yam, R.~Pogge,
K.~Sellgren, N.~Smith, V.~Stanishev, and the referee for comments. We
thank the SINGS Legacy Survey for making their data publicly available
and all the members of the LBT partnership who contributed to the
Science Demonstration Time observations. This work is based in part on
archival data obtained with the SST, which is operated by the JPL,
Caltech under a contract with NASA. This research has made use of NED,
which is operated by the JPL, Caltech, under contract with NASA and the
HEASARC Online Service, provided by NASA's GSFC. JLP and KZS are
supported by NSF grant AST-0707982, MDK by DOE grant DE-FG02-91ER40690,
and JFB and HY by NSF CAREER grant PHY-0547102.

\end{document}